\documentclass[sigconf]{acmart}

\usepackage[inline]{enumitem}

\AtBeginDocument{%
  }

\setcopyright{rightsretained}
\copyrightyear{2024}
\acmYear{2024}
\acmDOI{XXXXXXX.XXXXXXX}

\acmBooktitle{REDACTED}

\begin{document}

\title{Panmodal Information Interaction}
\author{Chirag Shah}
\orcid{0000-0002-3797-4293}
\email{chirags@uw.edu}
\affiliation{%
  \institution{University of Washington, Seattle, WA}
  \country{USA}  
}

\author{Ryen W. White}
\orcid{0000-0002-3797-4293}
\email{ryenw@microsoft.com}
\affiliation{%
  \institution{Microsoft Research, Redmond, WA}
  \country{USA}
}

\renewcommand{\shortauthors}{Shah \& White}

\begin{abstract}
The emergence of generative artificial intelligence (GenAI) is transforming information interaction.
For decades, search engines such as Google and Bing have been the primary means of locating relevant information for the general population. They have provided search results in the same standard format (the so-called ``10 blue links''). 
The recent ability to chat via natural language with AI-based agents and have GenAI automatically synthesize answers in real-time (grounded in top-ranked results) is changing how people interact with and consume information at massive scale. These two \emph{information interaction modalities} (traditional search and AI-powered chat) coexist in current search engines, either loosely coupled (e.g., as separate options/tabs) or tightly coupled (e.g., integrated as a chat answer embedded directly within a traditional search result page). We believe that the existence of these two different modalities, and potentially many others, is creating an opportunity to re-imagine the search experience, capitalize on the strengths of many modalities, and develop systems and strategies to support seamless flow between them.
We refer to these as \emph{panmodal} experiences.\footnote{An alternative nomenclature would be \emph{multimodal} experiences. Multimodal is a common term in human-computer interaction for the availability of multiple modalities, but is also mostly used in AI these days to refer to the use of different content types (text, images, video, audio, and so on) in foundation model training and inference. As we will discuss, we also believe that multimodal is a subset of panmodal.} Unlike monomodal experiences, where only one modality is available and/or used for the task at hand, panmodal experiences make multiple modalities available to users (multimodal), directly support transitions between modalities (crossmodal), and seamlessly combine modalities to tailor task assistance (transmodal). While our focus is search and chat, with learnings from insights from a survey of over 100 individuals who have recently performed common tasks on these two modalities, we also present a more general vision for the future of information interaction using multiple modalities and the emergent capabilities of GenAI.
\end{abstract}

\maketitle

\section{Interacting with information}
Information is essential for effective decision making and action. Information systems such as search engines facilitate rapid and comprehensive access to that information. An information interaction modality describes the particular mode of how people engage with such a system, including different interaction paradigms (e.g., query-response, multi-turn dialog, proactive suggestions), our focus here, but also different input mechanisms (e.g., text, speech, touch), and different device types. For decades, mainstream search interfaces in search engines have offered just one information interaction modality: query-response or simply \emph{search} (comprising text query, hyperlinked search results, click-through to landing pages, and iterative query refinement as needed to find relevant information). 
While search engines have been a primary source of information for consumers, emerging modalities fueled by GenAI models such as Gemini and GPT-4o, that can reason over text, audio, and vision inputs and generate text and multimedia outputs, can complement search engine capabilities and have created new possibilities for information interaction \cite{white2023navigating}. 

The \emph{chat} interface is an essential component of many GenAI-based systems and is another information interaction modality in its own right. Multi-turn dialog has long shown promise as a way to engage with information systems \cite{oddy1977information}, but is now going mainstream in support of complex tasks via progress in GenAI \cite{zamani2023conversational} and in GenAI-based systems such as ChatGPT.\footnote{\url{https://chat.openai.com/}} Search engines can now show GenAI answers directly on result pages---minimizing user effort in examining search results but also removing human control over answer generation \cite{shneiderman2022human}---and let users follow up via multi-turn conversation for clarification or to seek additional information. 

When only one modality is available and/or used (e.g., search only, chat only), we call it {\em monomodal information interaction}. When multiple modalities are used separately or collectively for accessing information, we refer to it as {\em panmodal information interaction}. Information systems with both search and chat functionality may only be the beginning. With the advent of GenAI, we foresee considerable growth opportunity in panmodal experiences. Computer science researchers and practitioners are well placed to help envisage and develop new panmodal experiences.

\section{Panmodal Information Interactions}
To help inform our viewpoint, we surveyed those who have used multiple modalities of information interactions---search and chat---in recent months. This provided insights on how people are starting to use multiple modalities, as well as what factors lead to successes and failures of these modalities individually and collectively. The survey included a few multiple choice questions and several questions that elicited open-ended text responses---all seeking to obtain personal accounts of using search and chat modalities for accomplishing tasks of a personal and professional nature.

We surveyed a few hundred crowdworkers. After data cleaning to remove bots and low-quality responses, we had 130 responses to analyze. Respondents' reported genders were male (60\%), female (39\%), and other (1\%). They were all from the US, English-speaking, and from various professions and educational backgrounds. Respondents reported using search and chat to accomplish a wide range of personal and professional tasks. 

\begin{figure}[htbp]
  \centering
  \includegraphics[width=1.0\linewidth]{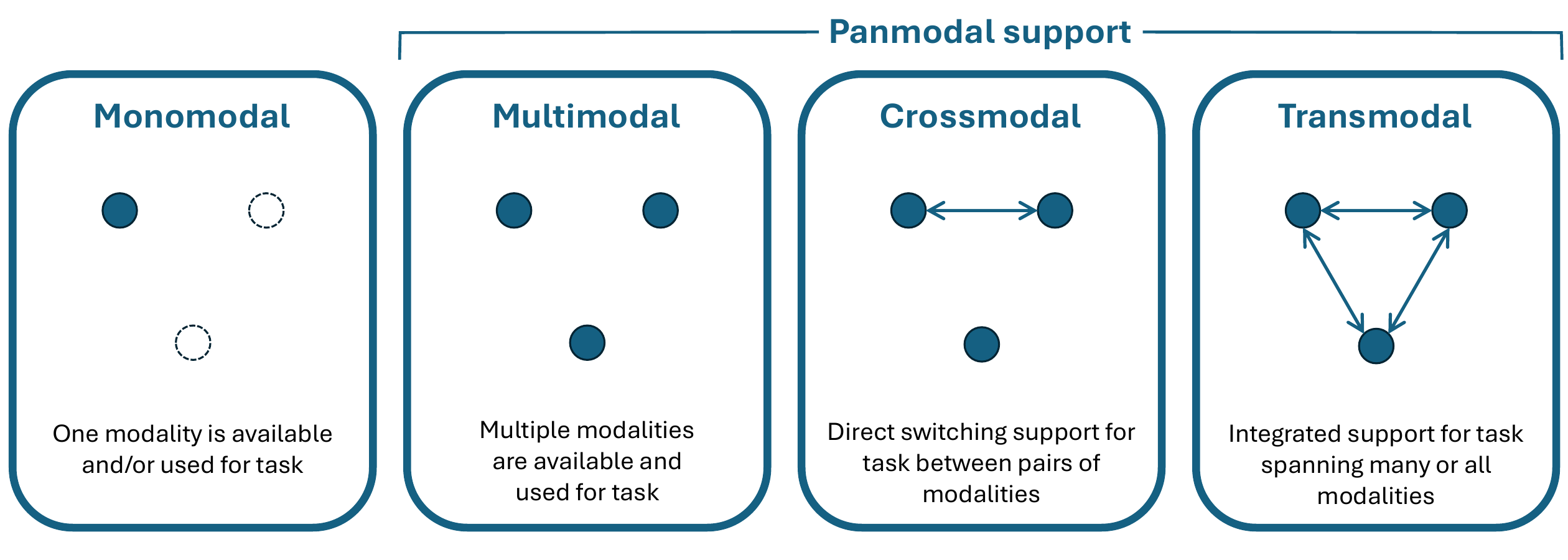}
  \caption{Examples of different modality usage patterns and panmodal support for information interaction.}
  \label{modality_types}
\end{figure}

We examined the survey responses with the lens of monomodal and panmodal support as shown in Figure \ref{modality_types}.

\subsection{Monomodal Information Interactions}
\begin{figure}[htbp]
  \centering
  \includegraphics[width=1.0\linewidth]{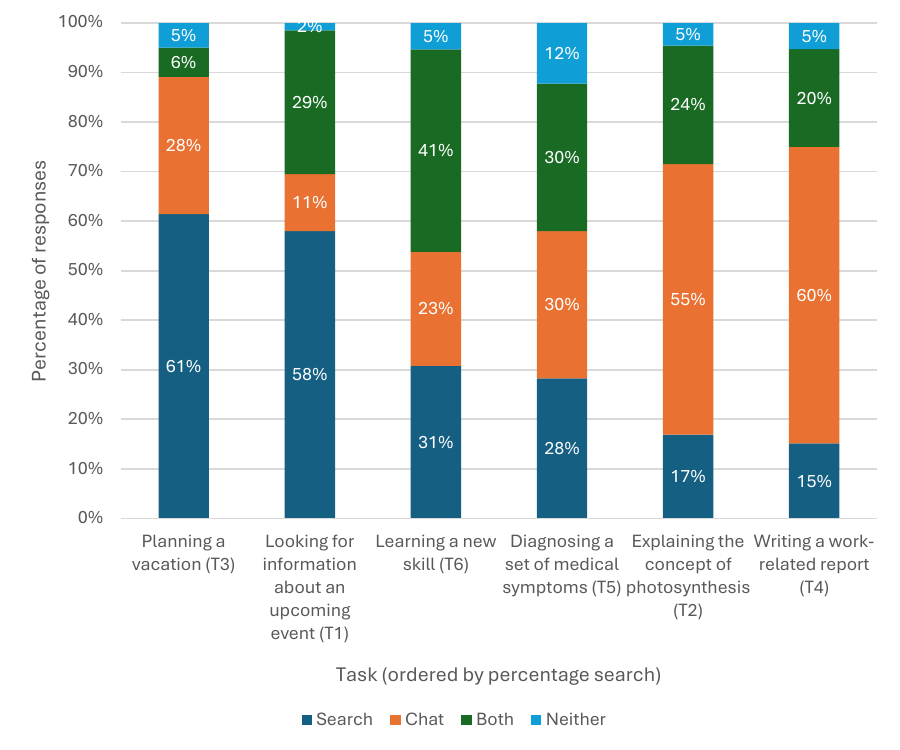}
  \caption{Distribution of modalities selected by survey respondents for various task examples. Tasks are ordered from left to right by the percentage of responses where search is the preferred modality.}
  \label{fig:tasks}
\end{figure}

In monomodal interactions, people select just one modality for the task.
To understand how they choose from different modalities available to them, we gave our respondents six specific tasks (see Figure \ref{fig:tasks}) and asked that they select the modality between search and chat that they find most appropriate. As shown in Figure \ref{fig:tasks}, people overwhelmingly pick search engines for informational and planning tasks (T1 and T3), while they pick chat-based modalities for understanding/learning and creation tasks (T2 and T4). They also prefer chat-based tools for learning tasks (T6), but search is still an important modality for such tasks as well as medical diagnosis (T5).

Respondents reported that they preferred traditional search engines when they wanted more options, details, or more diversity/comprehensiveness in their information. 
The aspect of familiarity also plays a big role in users choosing search engines over other modalities for their information needs. 
Conversely, users find chat-based systems more effective when they need direct and quick answers to specific questions, without having to sift through unrelated information. Users also prefer chat-based modalities for tasks that require learning, synthesis, or creation. 

It appears that the characteristics for picking a specific modality are tied to users' general preferences, but beyond that, we wanted to understand how they are using access to multiple modalities to accomplish real tasks.

\subsection{Multimodal Information Interactions}
Multimodal interfaces have been studied for decades in the human factors community \cite{oviatt2007multimodal}, including for information interactions. 
Search systems have long provided different facilities and search strategies based on user needs and goals \cite{croft1987i3r} and even suggested specific search engines given the current query \cite{white2008enhancing}.
 
As mentioned earlier, search engines now offer multiple modalities, search and chat, juxtaposed in the search interface. 
This is available on Bing and in Google as the so-called search generative experience (SGE), recently promoted by Google as {\em AI Overview}.\footnote{\url{https://blog.google/products/search/generative-ai-google-search-may-2024/}} We found that search is a preferred mode for information interaction for many personal tasks, such as finding recipes, planning a vacation, and shopping. Conversely, the chat-based modality dominates professional tasks, such as report generation, coding, and researching relevant topics for a business. The reasons we found through our survey shed some light on how users pick each modality and what they should watch for when deciding (Table \ref{reasons_challenges_table}).

For users, making informed selections about which modality to use also depends on sound mental models of system capabilities. People have developed mental models of search engines over decades of use. For chat-based systems, users' mental models are still being formed and users may experiment with different prompt formulations to tailor the system output to meet their expectations.
Showing both GenAI-powered chat experiences and traditional search results in a single result page, as in SGE, is also imprecise and can be overwhelming for users, potentially impeding task progress.

\begin{table*}[htbp]
\caption{Summary of reasons to use and challenges in using the two primary modalities (search and chat) offered by current commercial search systems, as derived from qualitative responses to survey questions.} 
\label{reasons_challenges_table}
\centering
\small
\begin{tabular}{|l|l|} % 'l' for left alignment
\hline
\multicolumn{2}{|c|}{\textbf{Search modality}} \\ % This will make the cell span two columns
\hline
\parbox[t]{7cm}{ % Adjust the width of the parbox as needed
\textbf{USE THIS MODALITY} when need for:
\begin{enumerate}[leftmargin=*]
\item \textbf{Broad and Unprompted Information:} Search engines often provide related or historical data that was not explicitly asked for. 
\item \textbf{Trustworthy and Verified Sources:} For academic or scholarly research, search engines provide direct access to authentic academic journals and books.
\item \textbf{Product Research and Shopping:} When deciding on a purchase, search engines provide detailed reviews, user experiences, specs, and multimedia content like images and videos. 
\item \textbf{Multiple Sources and Opinions:} Search engines are preferred to see multiple opinion-based pieces or reviews on a possible purchase or topic.
\item \textbf{Real-Time, Current, and Location-Based Information:} For real-time data like movie times, weather, game scores, or election results, search engines are more effective as they can provide up-to-date information. 
\item \textbf{Specific and Quick Information:} For quick and simple information, like restaurant details, holiday dates, or trivia, search engine are fast. 
\end{enumerate}
} & 
\parbox[t]{7cm}{ % Adjust the width of the parbox as needed
\textbf{CHALLENGES} with using modality include:
\begin{enumerate}[leftmargin=*]
\item \textbf{Irrelevant Results and Overwhelming Information:} Search engines could have results that are unrelated to specific queries or could overwhelm the user  with too much information, irrelevant results, or excessive advertisements.
\item \textbf{Inadequate Detail or Specificity:} Search engines may not be good for a need of detailed or highly specific information.
\item \textbf{Lack of Interactivity, Personalization, and Creativity:} Search engines could be inadequate in cases when an interactive experience is needed, such as asking follow-up questions or having a real-time discussion to solve a problem. 
\item \textbf{Difficulty in Finding Specific Items, Services, or Paywalled Content:} Search engines can be inadequate when users have difficulty finding specific items, services, or parts, or when they are unable to access paywalled or restricted content.
\item \textbf{Biased or Misleading Information:} Search engines may provide information that is biased, misleading, or inconsistent. 
\end{enumerate}
} % End of parbox
\\
\hline
\multicolumn{2}{|c|}{\textbf{Chat modality}} \\ % This will make the cell span two columns
\hline
\parbox[t]{7cm}{ % Adjust the width of the parbox as needed
\textbf{USE THIS MODALITY} when need for:
\begin{enumerate}[leftmargin=*]
\item \textbf{Direct and Quick Answers:} Chat-based systems are more efficient for direct and quick answers to specific questions, without having to sift through unrelated information. 
\item \textbf{Complex Queries and Specialized Knowledge:} For high-level topics or complex queries that require specialized knowledge, chat-based systems can be helpful in summarizing and clarifying points. 
\item \textbf{Personalized Assistance and Decision Making:} Chat-based systems can provide dynamic and personalized assistance provided when making decisions or seeking advice tailored to their specific situations. 
\item \textbf{Interactive Troubleshooting and Real-Time Engagement:} When dealing with technical issues or problems with devices, chat-based systems can be more effective in offering solutions and troubleshooting due to real-time interactions with ability to do back-and-forth. 
\item \textbf{Creative and Original Content:} Chat-based systems are superior for help with creative tasks such as brainstorming ideas, writing essays, emails, reviews, or creating campaigns. 
\item \textbf{Learning and Understanding:} Chat-based systems are beneficial for learning new concepts, languages, or understanding complex topics. 
\end{enumerate}
} & 
\parbox[t]{7cm}{ % Adjust the width of the parbox as needed
\textbf{CHALLENGES} with using modality include:
\begin{enumerate}[leftmargin=*]
\item \textbf{Difficulty Following Instructions, Handling Complex Queries, and Providing In-Depth Information:} This includes instances where the chat-based system fails to adhere to specific instructions given by the user, leading to outputs that do not meet the user’s requirements. 
\item \textbf{Limited Creativity, Originality, Personalization, and Handling of Sensitive Topics:} This category encompasses scenarios where the chat-based system fails to generate creative, original, or personalized content. 
\item \textbf{Inadequate for Quick Searches, General Information, and Real-Time Updates:} This category includes scenarios where chat-based systems could be slower or more complicated than traditional search engines for quick and simple queries. 
\item \textbf{Inadequate for Specialized Support and Complex Scenarios:} This category includes scenarios where the chat-based system fails to provide specialized support for technical issues, complex software applications, or specific customer service needs. 
\end{enumerate}
} \\ % End of parbox
\hline
\end{tabular}
\end{table*}

\subsection{Crossmodal Information Interactions}
While users have preferences for a modality given a task or their prior experiences, there are times when their initial choice does not pan out, or they may simply require some of the capabilities of a different modality, and they need to switch. We refer to these as \emph{crossmodal information interactions} and we investigated when and why users switch. Crossmodal has been explored in the context of search (e.g., text and voice-based input/output, different device types \cite{montanez2014cross}), but transitions between search and chat are less well understood and supported by current information systems.

From our survey responses, we learned that users switch from search to chat modality when what they are getting through search is not sufficiently interactive or tailored to their needs. They also found switching from search to chat beneficial when they needed detailed or highly specific information. 
In contrast, users switch from the chat to the search modality when they feel that search will give them a more comprehensive and diverse set of results than what they received through chat.

\begin{figure}[t!]
  \centering
 \includegraphics[width=1.0\linewidth]{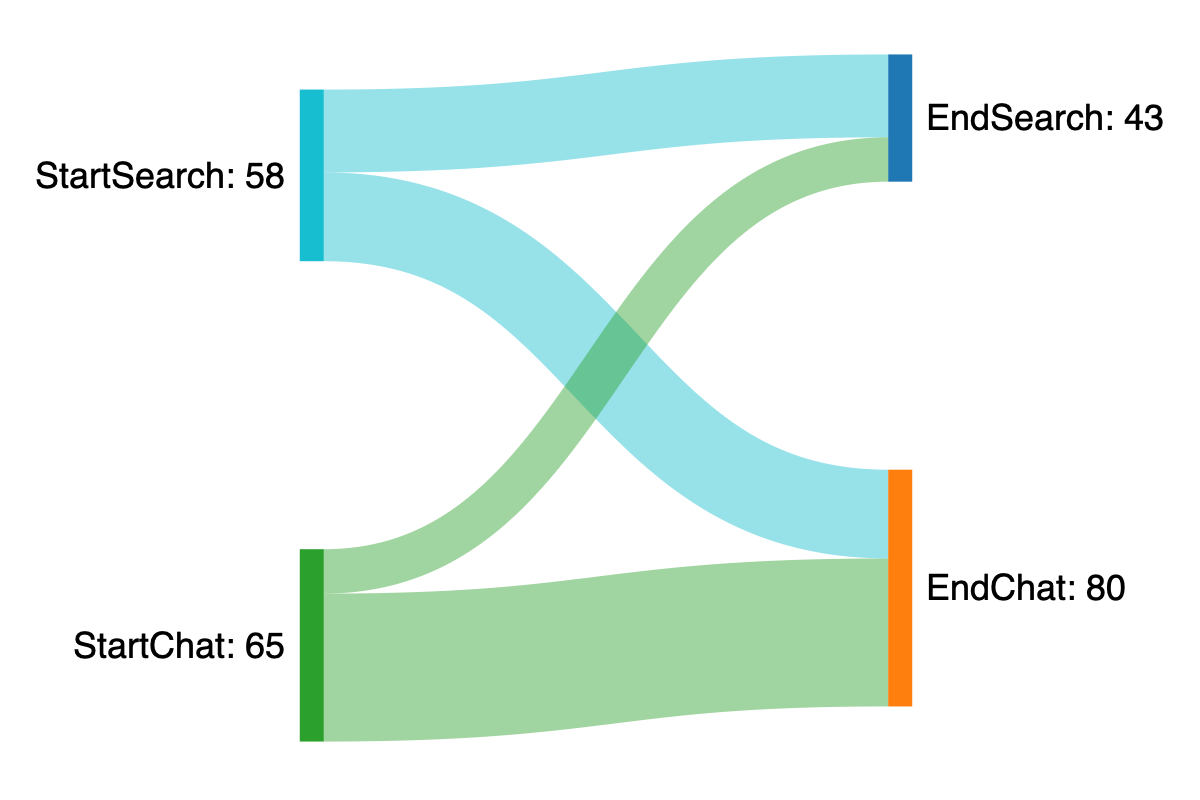}
 \caption{Evidence of crossmodality information interactions as people move from one modality to another to accomplish a task. The flow from search to chat is much greater than chat to search. The numbers are the number of respondents in each start and end state. Note that a small number of ambiguous responses (n=7) were removed from the data for this analysis.}
 \label{fig:crossmodal}
 \end{figure}

Figure \ref{fig:crossmodal} shows how often our respondents switch from one modality to the other. As we can see, when users start with search, they were almost equally likely to continue with search and switch to chat. On the other hand, when they start with chat, they are a lot less likely to switch to search. This disparity in crossmodality can be attributed to some of the strengths of the chat-based modality, including its ability to provide more direct answers that eliminate the need to explore further and the interactive nature of the modality that allows the users to more effectively reshape their requests to achieve desired outcomes.

We note that current mainstream search experiences are multimodal with very limited crossmodal support (e.g., GenAI-powered chat on result pages).
An information system could engage with users to help with this switching decision and suggest the most suitable modality for the current task or even task step/stage. 
Switching considerations also extend beyond the purview of technological solutions, e.g., are simply about modality awareness, which can alter usage habits, and connected to time spent with the chat modality, in order to learn how best to use its functionality.

\subsection{Transmodal Information Interactions}
As has been clear thus far, there is no one modality of information interactions that could satisfy users for the full range of their informational needs. More importantly, using multiple modalities strategically could help users not only accomplish their tasks more effectively, but also unlocks possibilities for tasks that are not traditionally performed through monomodal, multimodal, or even crossmodal approaches. There were a few such self-reported incidents of what we call \emph{transmodal information interactions} from our survey respondents.

Several respondents reported complex information needs that require more dynamic support (e.g., for using several different modalities for the same task, either sequentially \cite{montanez2014cross} or concurrently \cite{white2019multi}; representing and preserving task state; using GenAI to decompose tasks, plan actions, select and sequence modalities, and so on). These needs were often related to tasks such as healthcare (e.g., diagnosis and mitigation), legal (e.g., seeking legal advice), or research (e.g., creating summary reports for a business need). 
For example, while describing a report writing task, one respondent mentioned how a search engine could find examples of good reports while a chat-based system could generate a draft of the report, helping the user with both learning about and accomplishing a task better than before.

However, given that most current interfaces lack a clear support for running concurrent modalities with seamless integration and focus on the task at hand, our respondents were doing such transmodal work through ad hoc workflows. Considering the importance of supporting transmodal information interactions and current lack of such support, there is a dire need for platforms to not only facilitate the use of multiple modalities, but also to create user experiences with the most appropriate constellation of modalities and connections between them.
  
\section{Panmodal Futures}
Going forward, we expect that panmodality will play more of a central role in information interaction. There are clear benefits to developing specialized modalities with specific capabilities, in their complimentarity, and in efforts to bring them together as needed for tasks. Future information systems may well be built around different modalities in a unified, seamless way, with accommodations for adapting system operation and the user experience when additional personal/contextual information is available. This can be done within a single application or by using a federated approach, surfacing modalities provided by a range of different applications, including those developed by third parties.

Panmodal experiences will utilize different information interaction modalities depending on user preferences and/or the nature of the search task. Users can focus on tasks and the system selects or suggests the most appropriate modality or modality sequence. This requires task modeling and new routing mechanisms capable of matching tasks to relevant modalities. Improvements to personalization, intent understanding, and context awareness will yield more accurate and appropriate modality selection. 
For example, a user could start a conversation via the audio interface in an automobile, continue that with touch interface on their smartphone, and then explore the options through gestures on a wall display -- all driven by human initiative and/or guided by a GenAI agent. The modalities could be served by different providers and the selections and transitions could be brokered by agents engaging with users and/or other agents \cite{wu2023autogen}, e.g., to provide diverse perspectives on the available options. 

There are additional information interaction modalities beyond search and chat, e.g., bespoke interfaces generated natively by GenAI for the task at hand, interactive visualizations generated by GenAI akin to dynamic queries \cite{ahlberg1992dynamic}, proactive recommendations from GenAI based on audio and vision sensing, and moving from information access to information use, in GenAI-based operation or suggestion of tools to support task completion \cite{schick2024toolformer}. This will also expand beyond interaction paradigms into new modes of interaction (e.g., tactition, gesture, eye gaze) and new device types (e.g., smart rings, smart glasses), where GenAI could help interpret signals and add intelligence.
Determining the most appropriate type of panmodality support could serve as useful reminders to users even if we do not directly support their use in-situ in the application or support transitions to them in other applications. For example, for a task where transmodal support is best suited, GenAI could enumerate potential modalities and candidate sequence orders per a generated plan. This could also be used as a way to develop user awareness of different modalities available to them and help evolve their usage habits over time to make better use of those options on their own. 

There are clearly many challenges that need to be considered and addressed in developing panmodal experiences (e.g., representing tasks across modalities, understanding which modalities are available and applicable for the current task, communicating relevant modality affordances to users, making transitions maximally seamless). As with all applications of AI at scale, there are important practical considerations including safety, security, and privacy (e.g., what personal data (if any) is used for modality suggestion, how is personal data shared between modalities). We must pay close attention to these types of issues during the design of panmodal experiences and not view the issues as an afterthought. Only by prioritizing user trust can we build ``better together'' panmodal experiences that unite modalities safely and securely to help us all tackle and complete our information tasks more effectively.

\balance
\bibliographystyle{ACM-Reference-Format}
\bibliography{sample-base}

\end{document}